\documentclass[letterpaper]{article} 
\usepackage{aaai2026}  
\usepackage{times}  
\usepackage{helvet}  
\usepackage{courier}  
\usepackage[hyphens]{url}  
\usepackage{graphicx} 
\usepackage{natbib}  
\usepackage{caption} 
\usepackage{algorithm}
\usepackage{algorithmic}
\usepackage{multirow}
\usepackage{pifont}
\usepackage{subfigure}
\usepackage{textcomp}
\usepackage{newfloat}
\usepackage{listings}
\DeclareCaptionStyle{ruled}{labelfont=normalfont,labelsep=colon,strut=off} 
\floatstyle{ruled}
\newfloat{listing}{tb}{lst}{}
\floatname{listing}{Listing}

\title{CO2-Meter: A Comprehensive Carbon Footprint\\Estimator for LLMs on Edge Devices}
\author{
    Zhenxiao Fu,
    Chen Fan,
    Lei Jiang
}
\affiliations{
Indiana University Bloomington\\
\{zhfu,fc7,jiang60\}@iu.edu
}
\begin{document}

\maketitle

\begin{abstract}
LLMs have transformed NLP, yet deploying them on edge devices poses great carbon challenges. Prior estimators remain incomplete, neglecting peripheral energy use, distinct prefill/decode behaviors, and SoC design complexity. This paper presents \textit{CO2-Meter}, a unified framework for estimating operational and embodied carbon in LLM edge inference. Contributions include: (1) equation-based peripheral energy models and datasets; (2) a GNN-based predictor with phase-specific LLM energy data; (3) a unit-level embodied carbon model for SoC bottleneck analysis; and (4) validation showing superior accuracy over prior methods. Case studies show \textit{CO2-Meter}'s effectiveness in identifying carbon hotspots and guiding sustainable LLM design on edge platforms. Source code: \url{https://github.com/fuzhenxiao/CO2-Meter}.
\end{abstract}

\section{Introduction}
\label{s:intro}

LLMs~\cite{qwen} now reach human-level performance on diverse NLP tasks, enabled by large transformers, extensive training, and massive pre-training corpora. While once cloud-only, privacy and QoS concerns~\cite{Adekanye:AAAI2024} are pushing inference to edge devices, powering applications from autonomous driving~\cite{Adekanye:AAAI2024} and VR assistants~\cite{Min:ISMAR2024} to human-robot interaction~\cite{Kim:HCI2024} and healthcare robots~\cite{venkataswamy2024realization}. This shift could sharply raise emissions: ARM projects 40\% annual growth in edge devices through 2035~\cite{sparks2017route}, expanding \textit{operational} footprints from usage and \textit{embodied} footprints from manufacturing~\cite{Gupta:ISCA2022}. LLMs exacerbate both—high inference costs increase operational emissions, while demand for NPUs~\cite{Song:ISSCC2019}, GPUs, and memory boosts embodied emissions. Without intervention, edge-device emissions may surpass global data centers by 2028~\cite{sparks2017route}, highlighting the urgency of measuring LLM carbon footprints on edge platforms.

Previous work lacks a comprehensive carbon footprint modeling tool for LLM inference on edge devices, overlooking both operational and embodied carbon emissions:
\begin{itemize}
\item \textit{Operational Carbon}: Existing LLM carbon/energy models~\cite{faiz2024llmcarbon,Fu:ARXIV2024,luccioni2024power,Ukarande:ISLPED2024} often ignore peripheral energy costs—data acquisition (sensors, cameras, mics), transmission (WiFi, Bluetooth), and output (audio, display)—despite their importance for on-device LLMs. Prior work mainly profiles energy for LLM training~\cite{faiz2024llmcarbon} and inference~\cite{Fu:ARXIV2024,luccioni2024power,Ukarande:ISLPED2024} in the cloud, or for small CNNs on edge devices~\cite{Tu:SEC2024,Kasioulis:IC2E2024,Chen:ARXIV2024}. But LLM inference energy on constrained edge platforms is largely unstudied, and CNN-based estimators~\cite{Tu:SEC2024} fail to capture the distinct compute and memory demands of LLM prefill and decode phases.

\item \textit{Embodied Carbon}: Studies~\cite{Chen:ARXIV2024,Pirson:JCP2021} show non-computing parts (casings, PCBs, batteries) dominate embodied carbon in low-end IoT (Internet of Things) devices. LLMs, however, demand high-performance NPUs~\cite{ale2024empowering}, GPUs, and large memory, driving emissions higher. Cloud servers estimate embodied carbon by multiplying carbon per unit area by total chip area (CPUs, GPUs, DRAMs)~\cite{Gupta:ISCA2022,faiz2024llmcarbon}, but edge devices consolidate units into a single SoC—making chip-level models inadequate for capturing unit-level emissions and pinpointing embodied carbon bottlenecks.
\end{itemize}

To address the limitations in prior work, this paper presents \textit{CO2-Meter}, a comprehensive model for estimating the end-to-end carbon footprint of deploying LLMs on edge devices. Our contributions can be summarized as:
\begin{itemize}
\item \textit{Peripheral Operation Energy Models and Dataset}: We profile operational energy consumption from peripheral operations—data acquisition (e.g., cameras), transmission (e.g., WiFi, Bluetooth), and output (e.g., audio, display)—on edge devices, and construct a dataset. Equation-based models are proposed to estimate the energy of various peripheral operations on edge devices.

\item \textit{LLM Inference Energy Prediction and Dataset}: An LLM inference energy dataset is compiled from real-world request traces across multiple devices. A GNN-based predictor is presented to accurately predict operational energy consumption for the prefill and decode phases of an LLM inference under diverse configurations.

\item \textit{Unit-Level Embodied Carbon Modeling for SoCs}: We propose a unit-level embodied carbon model to assess the carbon overhead of critical computing units in edge SoCs. This model identifies embodied carbon bottlenecks, supporting efficient design and deployment of edge devices optimized for LLM inferences.

\item \textit{Model Validation \& Use Case Studies}: Extensive validation demonstrates the accuracy of our models and their superiority over previous approaches. Use case studies highlight CO2-Meter’s ability to pinpoint operational and embodied carbon hotspots, informing sustainable LLM deployment strategies on edge platforms.
\end{itemize}

\begin{figure}[t!]
\centering
\begin{minipage}{.6\linewidth}
\centering
\includegraphics[width=\linewidth]{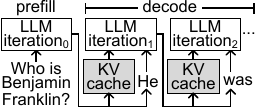}
\vspace{-0.2in}
\caption{Autoregressive inference.}
\label{f:co2_auto_infer}
\end{minipage}
\hfill
\begin{minipage}{.35\linewidth}
\centering
\includegraphics[width=\linewidth]{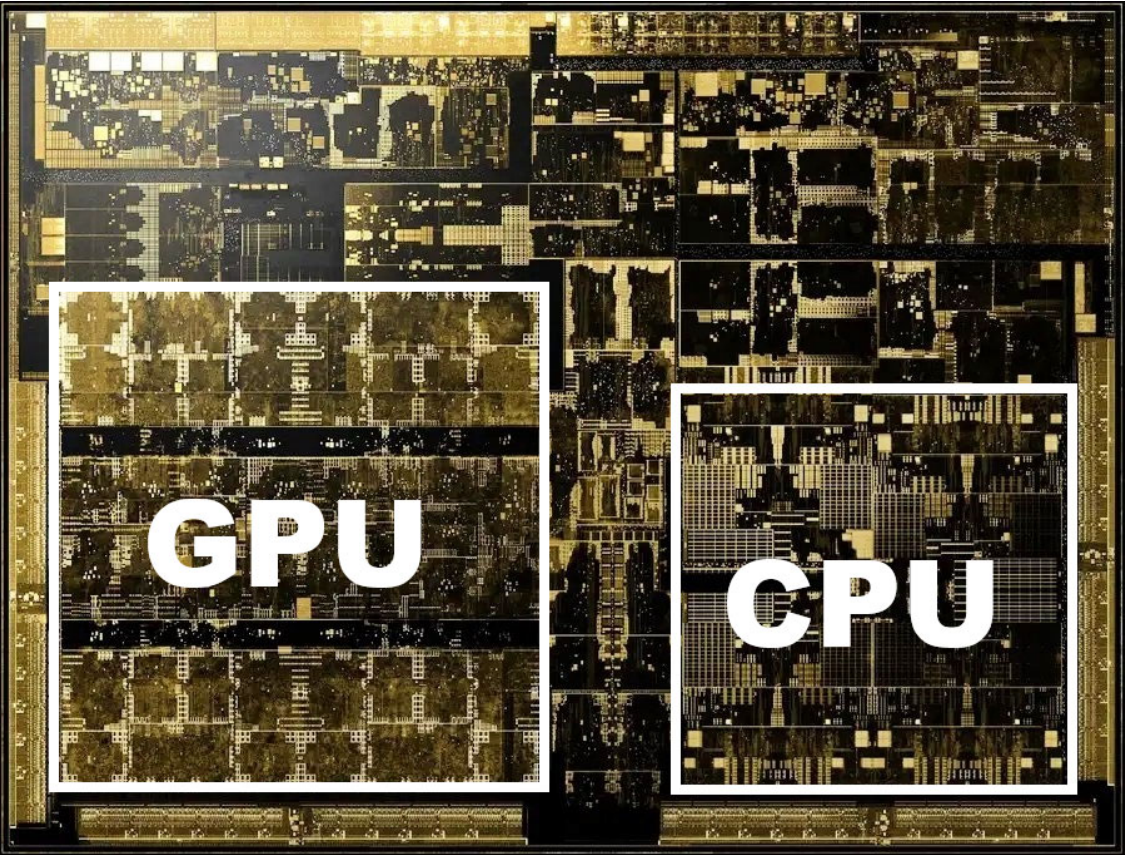}
\vspace{-0.2in}
\caption{A SoC die.}
\label{f:co2_soc_edge}
\end{minipage}
\vspace{-0.1in}
\end{figure}

\begin{figure}[t!]
\centering
\includegraphics[width=\linewidth]{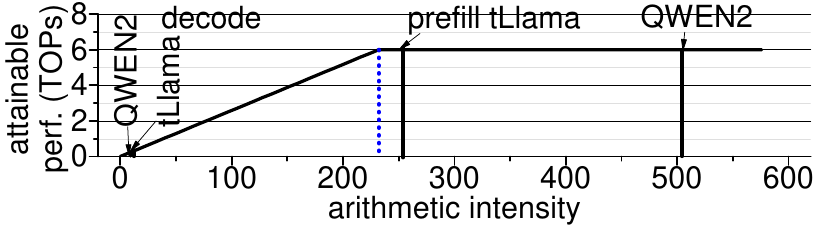}
\vspace{-0.2in}
\caption{The distinctive hardware characteristics of the prefill and decode phases in an LLM inference on rk3588.}
\label{f:co2_tech_moti}
\vspace{-0.2in}
\end{figure}

\section{Background and Motivation}

\subsection{Prior Operational Energy Estimators}
\textbf{Peripheral Operations on Edge Devices}. Most CNN~\cite{Tu:SEC2024,Kasioulis:IC2E2024,Chen:ARXIV2024} and LLM~\cite{Fu:ARXIV2024,luccioni2024power,Ukarande:ISLPED2024,faiz2024llmcarbon} energy estimators overlook peripheral energy use—data collection (sensors, cameras, mics), transmission (WiFi, Bluetooth), and output (audio, displays). Existing WiFi~\cite{Li:ICDCS2014} and Bluetooth~\cite{Negri:ICCNC2006} models target multi-device setups and are overly complex for single-edge devices, while models for key peripherals like cameras, microphones, speakers, and video output remain absent.

\textbf{LLM Inference}. LLM training~\cite{faiz2024llmcarbon} and inference~\cite{Fu:ARXIV2024,luccioni2024power,Ukarande:ISLPED2024} are well profiled in cloud settings, but edge studies largely focus on CNN latency~\cite{Zhang:MobSys2021,Liu:ICPP2023,Hu:EUROSYS2024,Yi:NIPS2023} with limited energy analysis~\cite{Tu:SEC2024,Kasioulis:IC2E2024,Chen:ARXIV2024}. No prior work models the energy cost of LLM autoregressive inference on edge. CNN-based estimators~\cite{Tu:SEC2024,Kasioulis:IC2E2024,Chen:ARXIV2024} treat inference as one phase, missing LLMs’ distinct compute-heavy prefill (parallel token processing) and memory-heavy decode (sequential KV-cache access) phases, as shown in Figures~\ref{f:co2_auto_infer} and~\ref{f:co2_tech_moti}.

\subsection{Limitations of Prior Embodied Carbon Models}

Unlike in the cloud, where operational emissions dominate~\cite{wu2022sustainable}, embodied carbon is often the main contributor on edge devices~\cite{Gupta:ISCA2022}. Studies~\cite{Chen:ARXIV2024,Pirson:JCP2021} link embodied emissions in low-end IoT devices to non-computing parts (casings, PCBs, batteries), which lack the NPUs/GPUs needed for LLMs~\cite{Suzen:HORA2020,Rockchip:Rockchip2024,NVIDIA:Jetson2024}. In cloud servers, discrete chips (CPUs, GPUs) are modeled by chip area~\cite{Gupta:ISCA2022,faiz2024llmcarbon}, but edge devices merge CPUs, GPUs, and NPUs into a single SoC (Figure~\ref{f:co2_soc_edge}). Chip-level models miss these SoC designs, limiting embodied carbon analysis for LLM inference on edge.

\subsection{Comparison with Prior Work}

Table~\ref{t:co2_related_work} compares prior work with CO2-Meter. Most studies address LLM carbon footprints in the cloud~\cite{faiz2024llmcarbon,Fu:ARXIV2024,luccioni2024power,Ukarande:ISLPED2024}, while edge research largely targets CNN/vision transformer latency~\cite{Zhang:MobSys2021,Liu:ICPP2023,Hu:EUROSYS2024,Yi:NIPS2023} with limited energy focus~\cite{Tu:SEC2024,Kasioulis:IC2E2024,Chen:ARXIV2024}. No work models LLM inference energy on edge or separates prefill and decode phases. Embodied carbon studies~\cite{Pirson:JCP2021} mostly cover non-computing parts in low-end IoT. CO2-Meter fills these gaps by jointly modeling operational and embodied carbon for autoregressive LLM inference, capturing core and peripheral operations, phase-specific energy, and unit-level SoC emissions.
\begin{table}[t!]
\setlength{\tabcolsep}{2pt}
\footnotesize
\centering
\caption{Comparing CO2-Meter against Prior Works.}
\label{t:co2_related_work}
\begin{tabular}{|c||c|c|c|c||c|}
\hline
\multirow{3}{*}{scheme} & \multicolumn{4}{c||}{operational carbon}        & unit-level \\\cline{2-5}
                        & edge  & energy    & autoregressive      & peripheral & embodied\\
                        & focus & profiling & inference           & energy     & carbon    \\\hline
1, 2, 3, 4   & \ding{55} & \ding{51}  & \ding{51} & \ding{55} & \ding{55}  \\\hline
5, 6, 7, 8   & \ding{51} & \ding{55}  & \ding{55} & \ding{55} & \ding{55} \\\hline
9, 10, 11    & \ding{51} & \ding{51}  & \ding{55} & \ding{55}  & \ding{55}  \\\hline
12           & \ding{51} & \ding{55}  & \ding{55} & \ding{55}  & \ding{55}  \\\hline
\textbf{CO2-Meter} & \ding{51} & \ding{51} & \ding{51} & \ding{51}  & \ding{51}  \\\hline
\end{tabular}

\vspace{0.05in}
\scriptsize{\textbf{Note:} 
1 = \cite{faiz2024llmcarbon};
2 = \cite{Fu:ARXIV2024};
3 = \cite{luccioni2024power};
4 = \cite{Ukarande:ISLPED2024};
5 = \cite{Zhang:MobSys2021};
6 = \cite{Liu:ICPP2023};
7 = \cite{Hu:EUROSYS2024};
8 = \cite{Yi:NIPS2023};
9 = \cite{Tu:SEC2024};
10 = \cite{Kasioulis:IC2E2024};
11 = \cite{Chen:ARXIV2024};
12 = \cite{Pirson:JCP2021}.}
\vspace{-0.2in}
\end{table}

\section{CO2-Meter}
\label{s:co2meter}

To estimate the end-to-end carbon footprint of LLM inference on edge devices, CO2-Meter separates the analysis into operational and embodied carbon modeling. Operational carbon is computed by estimating the energy consumption of both peripheral operations and LLM inferences, scaled by the carbon intensity of the edge device location (kgCO\(_2\)/kWh)~\cite{Gupta:ISCA2022}. The energy consumption of various peripheral operations such as data sensing, transmission, and output is modeled by equation-based approaches, while the energy consumption of LLM inferences is estimated using a GNN model. Embodied carbon is quantified by modeling the contributions of individual computing units within integrated edge SoCs.

\begin{figure*}[t!]
\centering
\subfigure[Energy dataset.]{
   \includegraphics[width=1.2in]{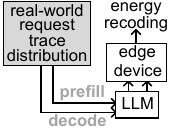}
   \label{f:co2_real_trace}
}
\hspace{-0.1in}
\subfigure[Graph representation.]{
   \includegraphics[width=2in]{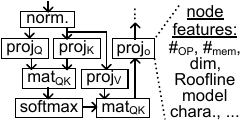}
   \label{f:co2_graph_nn}
}
\hspace{-0.1in}
\subfigure[2-phase energy predictor.]{
   \includegraphics[width=3in]{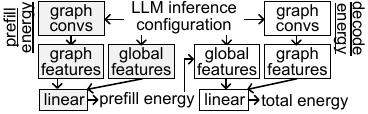}
   \label{f:co2_meter_gnn}
}
\vspace{-0.1in}
\caption{The LLM inference operational energy predictor of CO2-Meter.}
\vspace{-0.2in}
\end{figure*}

\subsection{Operational Energy Modeling}

\subsubsection{Peripheral Operation}  
This section models the operational energy and power consumption of peripheral operations. For certain operations, only power models are provided, with energy consumption computed as the product of power and execution time.
\begin{itemize}
\item \textit{Networking (WiFi/Bluetooth)}: Simplifying existing complex power models for WiFi~\cite{Li:ICDCS2014} and Bluetooth~\cite{Negri:ICCNC2006}, we propose a straightforward energy model:  
\begin{equation}
E_{net} = P_{static} \cdot t + E_{bit} \cdot S_{data},
\label{e:co2_comm_energy}
\end{equation}  
where \(E_{net}\) is the network energy, \(P_{static}\) represents the static power of the network interface (including chips and antenna), \(t\) is the transmission time, \(E_{bit}\) denotes the energy per bit of data transfer, and \(S_{data}\) is the transferred data size.

\item \textit{Camera}: The energy consumption for camera operations is modeled as:  
\begin{equation}
E_{cam} = P_{static} \cdot t + E_{frame} \cdot num_{frame},
\label{e:co2_cam_energy}
\end{equation}  
where \(E_{cam}\) is the total energy for camera usage, \(P_{static}\) denotes the static power of the camera interface (including image-capturing chips and lens), \(t\) is the duration of camera usage, \(E_{frame}\) indicates the energy per frame, and \(num_{frame}\) is the total number of frames captured.

\item \textit{Microphone}: The energy consumption of microphone operations is modeled as:  
\begin{equation}
E_{mic} = P_{static} \cdot t + E_{sample} \cdot num_{sample},
\label{e:co2_mic_energy}
\end{equation}  
where \(E_{mic}\) is the total microphone energy, \(P_{static}\) is the static power (e.g., analog-to-digital converters~\cite{Kim:AS2022} and supporting circuits), \(t\) is the microphone usage duration, \(E_{sample}\) is the energy per sample, and \(num_{sample}\) is the total samples captured.

\item \textit{Video Output}: Modern edge SoCs utilize multimedia units to generate HDMI output signals. The power consumption of the multimedia unit during HDMI signal generation is modeled as:  
\begin{equation}
P_{video} = P_{static} + P_p \cdot N_p,
\label{e:co2_hdmi_energy}
\end{equation}  
where \(P_{video}\) represents the video signal generation power, \(P_{static}\) is the static power consumed by the multimedia unit, \(P_p\) is the power required per pixel, and \(N_p\) is the number of pixels.

\item \textit{Speaker}: The speaker power consumption is primarily used for membrane vibrations and modeled as:  
\begin{equation}
P_{spk} = \frac{1}{1 + \exp(\alpha \cdot V) + \beta},
\label{e:co2_spk_energy}
\end{equation}  
where \(P_{spk}\) is the total speaker power, \(\alpha\) and \(\beta\) are two fitting parameters, and \(V\) is the sound volume.

\item \textit{Display}: We used a TFT Liquid Crystal Display (LCD) as our monitor. We adopted a TFT LCD power model from~\cite{Cheng:ITCE2004}:
 \begin{equation}
  P_{TFT} = a + b\cdot x + c\cdot x^2
	\label{e:co2_lcd_energy}
 \end{equation}  
where \(P_{TFT}\) is the LCD power consumption, \(x\) is the pixel grey value $\in[0, 255]$, and $a$-$c$ are fitting parameters.

\item \textit{Image/Voice-to-Text Conversions}: In certain LLM applications, such as in-home healthcare systems~\cite{venkataswamy2024realization}, inputs and outputs may include images or voice instead of text. Image-to-text conversion is performed using OpenOCR~\cite{Du:2024}. Voice-to-text conversion is handled by RealtimeSTT~\cite{Beigel:2024}, while text-to-voice conversion is done via TTS~\cite{tts:2024}. The inference energy consumption of OpenOCR, RealtimeSTT, and TTS is estimated using a prior energy predictor~\cite{Tu:SEC2024} for small-scale CNNs.

\item \textit{System Background}: System background energy is calculated as the product of the SoC idle power and the application's execution time.
\end{itemize}

\subsubsection{LLM Autoregressive Inference}

To estimate LLM inference energy on edge devices, CO2-Meter employs a GNN model trained on a real-world LLM inference energy dataset. Given an LLM configuration and target edge device, the GNN predicts inference energy for unseen requests. Addressing limitations in prior CNN-based estimators~\cite{Tu:SEC2024,Kasioulis:IC2E2024,Chen:ARXIV2024}, our approach introduces the following innovations. We choose GNNs because their flexibility ensures that future variants of LLM architectures, each with potentially different graph structures, can all be accommodated by the same model.
\begin{itemize}
\item \textit{Energy Dataset from Real-World Traces}: Using LLM inference traces from Azure Cloud~\cite{MS:Azure2024} (Figure~\ref{f:co2_real_trace}), we build an energy dataset assuming user behavior is consistent across cloud and edge, differing only in execution site. Prefill and decode are treated separately: prefill measures energy for a sampled prompt plus one token, while decode records energy for the same prompt with a sampled output length.

\item \textit{Graph Representation}: As shown in Figure~\ref{f:co2_graph_nn}, our GNN models each transformer layer as a graph, with nodes as computational kernels and edges as data dependencies. Node features comprehensively encode kernel-level metrics. Aside from those included in Figure~\ref{f:co2_graph_nn},  arithmetic intensity,weight loads, activation loads/stores, KV cache loads/stores, and per-kernel inference time are also included. Edges capture the data flow between these kernels. The resulting graph is encoded using GNN layers such as GraphSAGE~\cite{Hamilton:NIPS2017}, graph attention networks~\cite{Velickovic:ICLR2018}, and graph isomorphism networks~\cite{xu2018how}.

\item \textit{Two-phase prediction}: For prefill energy (Figure~\ref{f:co2_meter_gnn}), the GNN extracts graph features from the LLM request, deriving global stats (op count, layer count, dimensions, memory, ...). These combine through a linear layer to estimate prefill energy. For total energy (prefill + decode), another GNN/linear layer pair is used, incorporating prefill energy into the global features.
\end{itemize}

\subsection{Embodied Carbon Footprint Modeling}

To compute the embodied carbon of each computing unit (e.g., CPU, GPU, or NPU) within an edge SoC, we propose a unit-level embodied carbon model:  
\begin{equation}
embodied\_carbon_{SoC} = \sum_{i=1}^n area_i \cdot CPA,
\label{e:co2_unit_embodied}
\end{equation}  
where \(area_i\) denotes the area of unit \(i\), \(CPA\) is the carbon emission per unit area~\cite{RE:PCB2024}, and \(n\) is the number of computing units. This model enables identification of the dominant contributors to the SoC total embodied carbon.

\begin{figure}[t!]
\centering
\includegraphics[width=\linewidth]{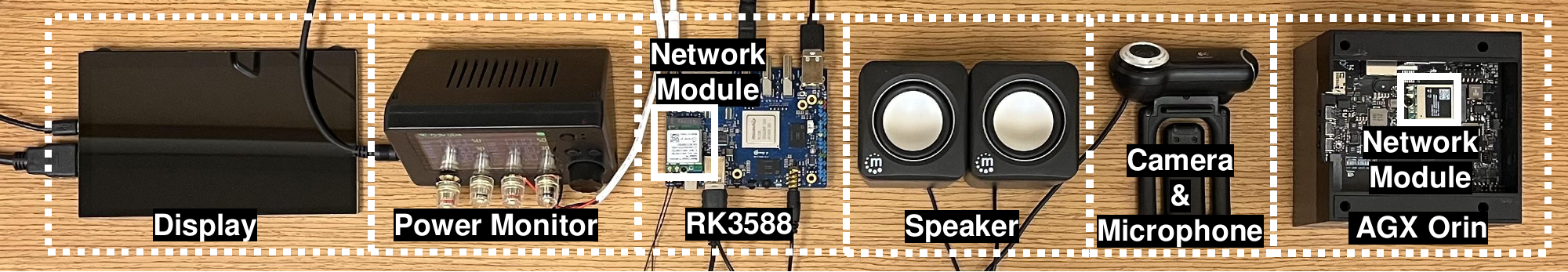}
\vspace{-0.2in}
\caption{The power measurement infrastructure.}
\label{f:co2_measure_devices}
\vspace{-0.2in}
\end{figure}

\section{Experimental Methodology}
\label{s:exp_method}

\textbf{Energy Measurement}. We develop a methodology for measuring the operational energy of peripherals and inference on edge devices (Figure~\ref{f:co2_measure_devices}). Inputs from WiFi, Bluetooth, cameras, or microphones are processed by the CPU, multimedia unit, NPU, or GPU before inference on the NPU/GPU, with outputs directed to displays or speakers. Energy consumption is recorded using an ODROID SmartPower3 meter~\cite{ODROID} at 200\,Hz. To ensure consistency and reliability across LLMs and devices, we follow:
\begin{itemize}
\item \textit{Operation-Specific Measurement}: Energy is computed as the difference in power between active and idle states for each operation, with all other conditions held constant.
\item \textit{Environmental Control}: Measurements are conducted at 25\,\textcelsius with a 10s cooldown between tests.
\end{itemize}

\begin{table}[t!]
\setlength{\tabcolsep}{3pt}
\footnotesize
\centering
\caption{The configuration of edge devices.}
\label{t:co2_hardware_device}
\vspace{-0.1in}
\begin{tabular}{|c||c|}
\hline
SoC       & Hardware Configuration \\\hline\hline
Rockchip  & CPU: 4 Cortex-A76 \& 4 A55; NPU: 6-TOPS Ethos\\
rk3588    & NPU; DRAM: 51.2GB/s 8GB 64-bit LPDDR5\\\hline

Rockchip  & CPU: 4 Cortex-A55; NPU: 1-TOPS Ethos\\
rk3568    & NPU; DRAM: 34.1GB/s 8GB 32-bit LPDDR4 \\\hline\hline

NVIDIA    & CPU: 12 Cortex-A78 v8.2; GPU: 275-TOPS CUDA\\
AGX Orin  & cores; DRAM: 204.8GB/s 32GB 256-bit LPDDR5\\\hline

NVIDIA    & CPU: 8 Cortex-A78 v8.2; GPU: 157-TOPS CUDA\\
Orin NX   & cores; DRAM: 102.4GB/s 16GB 128-bit LPDDR5 \\\hline

\end{tabular}
\vspace{-0.2in}
\end{table}

\textbf{LLMs.} We adopted the LLM-based virtual reality assistant application~\cite{Min:ISMAR2024} to generate Q\&A inference requests using selected lightweight LLMs, suitable for edge deployment due to resource constraints. 4 models ranging from 0.5 to 1.8 billion parameters were used: internlm2-chat-1.8b (INT)~\cite{cai2024internlm2}, qwen1.5-0.5b (Q1.5)~\cite{qwen}, tinyllama-1.1b-chat-v1.0 (LAM)~\cite{zhang2024tinyllama}, and qwen2-1.5b (Q2)~\cite{qwen}. Q2 was used specifically to evaluate CO2-Meter’s generalization to unseen LLM configurations. Prompt and output token length distributions were assumed to follow cloud-based inference patterns~\cite{MS:Azure2024}, with all requests executed at batch size 1.

\textbf{Edge and Peripheral Devices.} We evaluated Rockchip- and NVIDIA-based edge devices (Table~\ref{t:co2_hardware_device}). Rockchip devices use an NPU for LLM inference, while NVIDIA devices rely on a GPU. CO2-Meter was tested mainly on Rockchip rk3588 (rk)\cite{Rockchip:Rockchip2024} and NVIDIA AGX Orin (orin)\cite{NVIDIA:Jetson2024}, with rk3568 and Orin NX used to test generalization. Peripherals included a HAMTYSAN 7-inch $800\times480$ LCD, Manhattan 2600 speakers, and a Logitech QuickCam Pro 9000 webcam.

\textbf{Energy Dataset.} We collected energy data for $\sim200$ peripheral configurations to fit and validate equation-based energy models. For CO2-Meter’s GNN-based LLM inference energy predictor, we built a dataset of 40K measurements spanning LLMs (INT, Q1.5, LAM), request parameters, and SoCs (rk, orin), split into 32K/4K/4K for training/validation/testing. An extra 8K samples for Q2 on rk3568 and Orin NX evaluated generalization to unseen LLMs and hardware.

\textbf{Schemes.} To evaluate the accuracy of our equation-based peripheral energy models, we compared their predictions against real-world measurements. To assess the effectiveness of our GNN-based LLM inference energy predictor, we compared it against the following baselines:
\begin{itemize}
\item \textit{RF}: A random forest model~\cite{Zhang:MobSys2021} trained on global features such as total operations, transformer layer count, layer dimensions, and memory footprint.
\item \textit{NNLQP}: A GNN-based predictor~\cite{Liu:ICPP2023} using 2 GIN~\cite{xu2018how} layers, and trained on total inference energy without phase separation.
\item \textit{2P}: The sames as \textit{NNLQP}, except it is trained on a dataset separating LLM inference into prefill and decode phases.
\item \textit{CO2-Meter}: The sames as \textit{2P}, except it uses the GNN shown in Figure~\ref{f:co2_meter_gnn}.
\end{itemize}

\textbf{Setup.} All neural networks were implemented in PyTorch and trained using the Adam optimizer with a learning rate of 0.001 and a batch size of 32. Experiments were conducted on an NVIDIA A100 GPU.

\begin{figure}[t!]
\centering
\begin{minipage}{.47\linewidth}
\centering
\includegraphics[width=\linewidth]{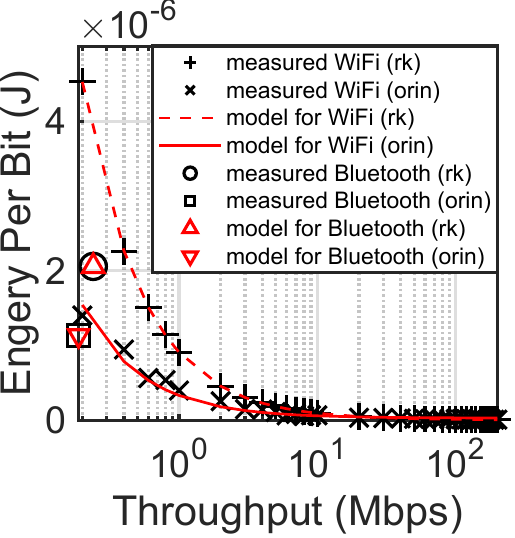}
\vspace{-0.2in}
\caption{Download val.}
\label{f:download}
\end{minipage}
\hfill
\begin{minipage}{.47\linewidth}
\includegraphics[width=\linewidth]{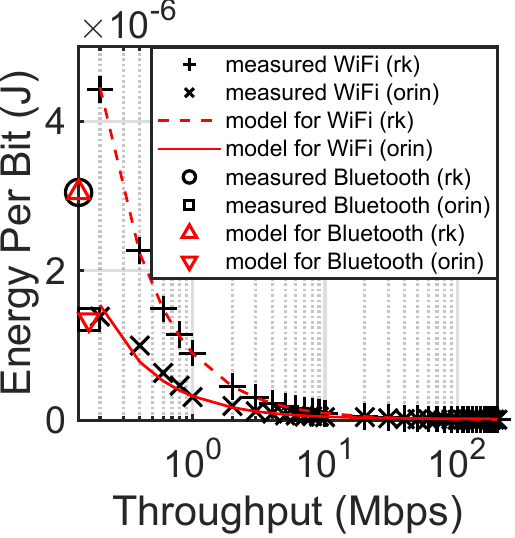}
\vspace{-0.2in}
\caption{Upload val.}
\label{f:upload}
\end{minipage}
\vspace{-0.15in}
\end{figure}

\begin{figure}[t!]
\centering
\begin{minipage}{.47\linewidth}
\includegraphics[width=\linewidth]{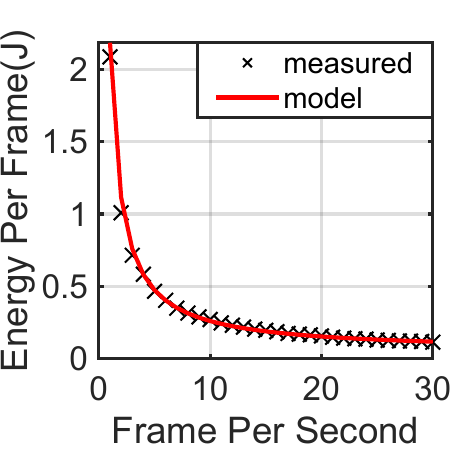}
\vspace{-0.2in}
\caption{Camera val.}
\label{f:cam}
\end{minipage}
\hfill
\begin{minipage}{.47\linewidth}
\includegraphics[width=\linewidth]{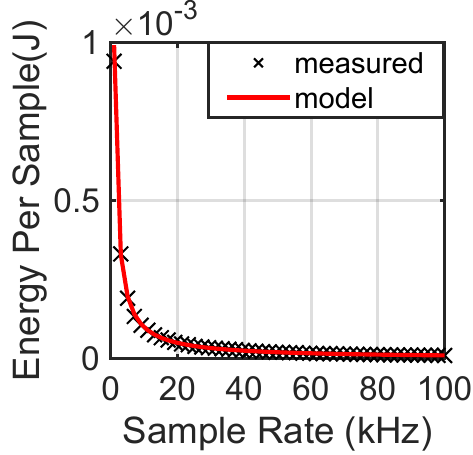}
\vspace{-0.2in}
\caption{Microphone val.}
\label{f:mic}
\end{minipage}
\vspace{-0.2in}
\end{figure}

\section{Validation}

\subsection{Peripheral Operation Energy Validation}
\textbf{WiFi \& Bluetooth.} We evaluated the download and upload energy models on rk3588 (rk) and AGX Orin (orin), as shown in Figures~\ref{f:download} and~\ref{f:upload}. Equation~\ref{e:co2_comm_energy} closely matches real-world measurements. For WiFi, energy per bit decreases exponentially with increasing bandwidth, as static power is amortized over a larger data volume. Orin, with more efficient antennas and support circuits, consistently outperforms rk in energy efficiency. The model achieves mean absolute errors of $2.72\times10^{-9}$\,J (rk) and $4.02\times10^{-8}$\,J (orin) for download, and $6.04\times10^{-9}$\,J (rk) and $3.42\times10^{-8}$\,J (orin) for upload, across bandwidths from 0 to 200 Mbps. For Bluetooth, which operates at a fixed bandwidth, the model yields mean absolute errors of $2.71/2.32\times10^{-8}$\,J (rk) and $4.26/4.85\times10^{-8}$\,J (orin) for download/upload, respectively.

\textbf{Camera \& Microphone.} Figures~\ref{f:cam} and~\ref{f:mic} validate our camera and microphone energy models. For the camera, energy per frame decreases with higher frame rates as static power is amortized across more frames. Equation~\ref{e:co2_cam_energy} yields a mean absolute error of $1.18\times10^{-2}$\,J. Similarly, the microphone model shows reduced energy per sample at higher sampling rates, with a mean absolute error of $1.80\times10^{-6}$\,J. These results demonstrate strong consistency with real-world measurements.

\begin{figure}[t!]
\centering
\begin{minipage}{.47\linewidth}
\includegraphics[width=\linewidth]{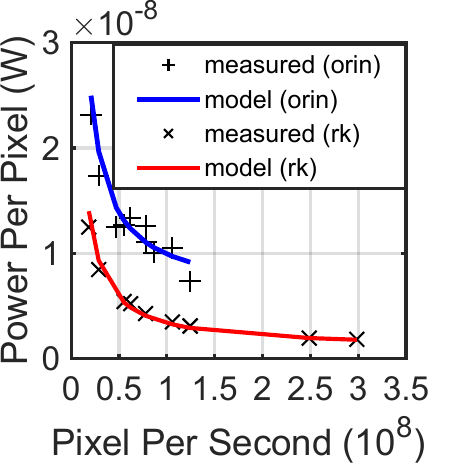}
\vspace{-0.2in}
\caption{Video output val.}
\label{f:gpu}
\end{minipage}
\hfill
\begin{minipage}{.47\linewidth}
\includegraphics[width=0.99\linewidth]{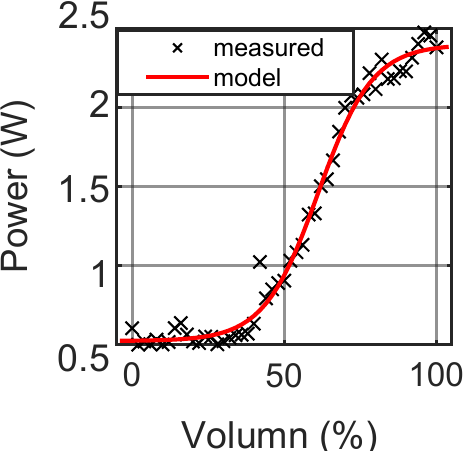}
\vspace{-0.2in}
\caption{Speaker val.}
\label{f:speaker}
\end{minipage}
\vspace{-0.1in}
\end{figure}

\begin{figure}[ht!]
\vspace{-0.05in}
\centering
\begin{minipage}{.47\linewidth}
\includegraphics[width=0.97\linewidth]{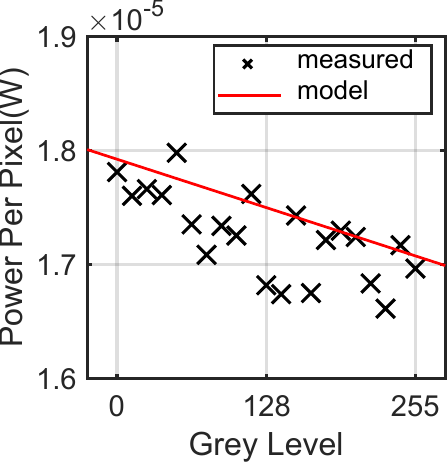}
\vspace{-0.1in}
\caption{TFT LCD val.}
\label{f:RGB}
\end{minipage}
\hfill
\begin{minipage}{.48\linewidth}
\includegraphics[width=0.92\linewidth]{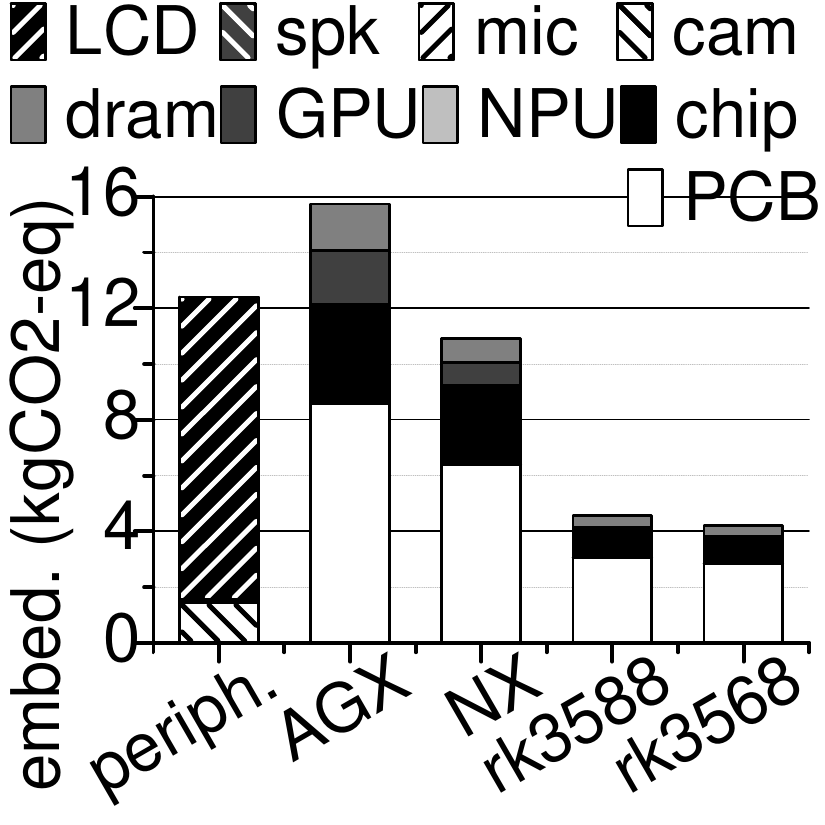}
\vspace{-0.1in}
\caption{Embod. carbon of edge \& periph. devices.}
\label{f:co2_embodied_all}
\end{minipage}
\vspace{-0.1in}
\end{figure}

\textbf{Video Output, Speaker, and Display.} We validated the power models for video output, speaker, and display, as shown in Figures~\ref{f:gpu}, \ref{f:speaker}, and~\ref{f:RGB}, respectively. For video output, rk3588 (rk) consumes less power than AGX Orin (orin) at the same resolution, due to its energy efficient multimedia unit. Power per pixel decreases with increasing pixel rate due to amortization of static power. The video output model (Equation~\ref{e:co2_hdmi_energy}) achieves mean absolute errors of $3.69\times10^{-10}$\,W (rk) and $1.21\times10^{-9}$\,W (orin). Speaker power increases nonlinearly with volume, and the speaker model (Equation~\ref{e:co2_spk_energy}) yields a mean absolute error of $5.39\times10^{-2}$\,W. The TFT LCD power model, fitted using three parameters, shows power reduction with increasing pixel gray level and achieves a mean absolute error of $4.71\times10^{-7}$\,W.

\textbf{Image/Voice-to-Text Conversions.} Following the methodology in~\cite{Tu:SEC2024}, we collected 14K energy samples for kernels in OpenOCR, RealtimeSTT, and TTS, partitioned into 10K for training, 1K for validation, and 3K for testing. A GNN-based energy predictor~\cite{Tu:SEC2024} was trained to model the energy consumption of image/voice-to-text and text-to-voice conversion tasks. On the test set, the predictor achieved 82\% accuracy within a 10\% deviation from actual values, consistent with previously reported results.

\textbf{Background Energy.} The model for background energy closely aligns with measured data, yielding negligible error.

\begin{table}[t!]
\centering
\setlength{\tabcolsep}{3pt}
\footnotesize
\caption{The MAPE comparison.}
\vspace{-0.1in}
\label{t:co2_mape_infer}
\begin{tabular}{|c|ccc|c|ccc|c|}\hline
\multirow{2}{*}{scheme} & \multicolumn{4}{c|}{rk3588 (\%)} & \multicolumn{4}{c|}{AGX Orin (\%)} \\ \cline{2-9}
             & INT  & Q1.5  & LAM  & \textbf{avg} & INT  & Q1.5 & LAM  & \textbf{avg} \\ \hline
RF           & 98.1 & 158.2 & 63.6 & 106.6        & 49.1 & 68.8 & 69.1 & 62.3 \\
NNLQP        & 23.4 & 26.4  & 49.4 & 33.1         & 29.8 & 29.7 & 24.9 & 28.1 \\
2P           & 11.9 & 11.4  & 12.3 & 11.8         & 16.3 & 22   & 17.9 & 18.7 \\
CO2-Meter    & 10.8 & 10.1  & 10   & \textbf{10.3}& 15.2 & 20.3 & 18.5 & \textbf{18} \\\hline
\end{tabular}
\vspace{-0.2in}
\end{table}

\subsection{LLM Inference Energy Validation}
\textbf{Seen Configuration.} We trained and evaluated the LLM inference energy predictor using data from LLMs (INT, Q1.5, LAM) on rk3588 and AGX Orin. Accuracy was measured by mean absolute percentage error (MAPE) and the share of predictions within 10\% error bounds (10\% EB)(Tables~\ref{t:co2_mape_infer},\ref{t:co2_b10_infer}). A $B\%$ at an $N\%$ bound means $B\%$ of predictions are within $N\%$ of ground truth. Results are stronger on rk3588 due to higher absolute energy values, which reduce relative error risk. While RF scores well under error bounds, its MAPE is high from large outliers. Phase-specific modeling in 2P boosts 10\% EB accuracy by 123–155\% over NNLQP. Our GNN further improves these bounds by 3\% on rk3588 and 19\% on AGX Orin, underscoring CO2-Meter’s advantage on high-end SoCs.

\begin{table}[t!]
\centering
\setlength{\tabcolsep}{3pt}
\footnotesize
\caption{The 10\% EB comparison.}
\vspace{-0.1in}
\label{t:co2_b10_infer}
\begin{tabular}{|c|ccc|c|ccc|c|}\hline
\multirow{2}{*}{scheme} & \multicolumn{4}{c|}{rk3588 (\%)} & \multicolumn{4}{c|}{AGX Orin (\%)} \\ \cline{2-9}
             & INT  & Q1.5 & LAM  & \textbf{avg}  & INT  & Q1.5 & LAM  & \textbf{avg} \\ \hline
RF           & 40.7 & 46.3 & 57.9 & 48.3          & 25.1 & 27   & 22   & 24.7 \\
NNLQP        & 34.2 & 42.5 & 15.2 & 30.6          & 11.8 & 21.3 & 15.3 & 16.1 \\
2P           & 67.9 & 69.2 & 67.4 & 68.2          & 48.3 & 29.5 & 45.1 & 41 \\
CO2-Meter    & 65.8 & 72.5 & 72.7 & \textbf{70.3} & 55.9 & 40.1 & 49.9 & \textbf{48.6} \\\hline
\end{tabular}
\vspace{-0.1in}
\end{table}

\begin{table}[t!]
\centering
\setlength{\tabcolsep}{3pt}
\footnotesize
\caption{The 10\% EB comparison on unseen configurations.}
\label{t:co2_unseen_inference}
\vspace{-0.05in}
\begin{tabular}{|c|ccc|ccc|}\hline
\multirow{2}{*}{LLM} & \multicolumn{3}{c|}{rk3568 (\%)} & \multicolumn{3}{c|}{Orin NX (\%)} \\ \cline{2-7}
                     & RF    & NNLQP  & CO2-Meter       & RF   & NNLQP  & CO2-Meter \\ \hline
Q2                   & 28.5  & 30.1   & \textbf{69.2}   & 17.5 & 25.7   & \textbf{45.1}\\\hline
\end{tabular}
\vspace{-0.1in}
\end{table}

\textbf{Unseen Configuration.} To assess generalization, we evaluated all schemes using 8K samples from Q2 inferences on rk3568 and AGX Orin—configurations not seen during training. Table~\ref{t:co2_unseen_inference} reports the 10\% EB comparison across methods. Accuracy drops significantly for RF under unseen LLM and SoC settings. Compared to NNLQP, CO2-Meter improves 10\% EB accuracy by 129\% on rk3568 and 75\% on AGX Orin, demonstrating its superior generalization to previously unseen configurations.

\begin{figure}[t!]
\centering
\includegraphics[width=\linewidth]{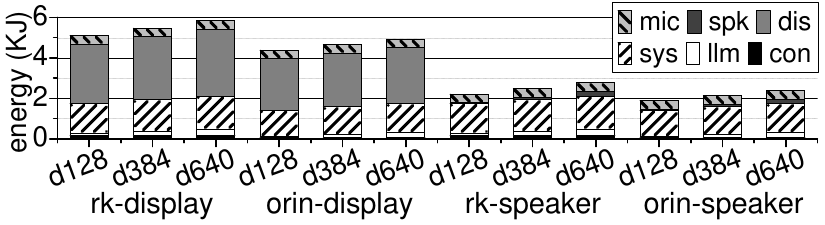}
\vspace{-0.2in}
\caption{The operational energy of the LLM-based virtual reality assistant with a microphone input (d$n$: generated token length $n$; mic: microphone; spk: speaker; dis: display; sys: system background; llm: LLM; con: conversion).}
\label{f:carbon_app_en1}
\vspace{-0.1in}
\end{figure}

\subsection{Embodied Carbon Validation and Calculation}
\label{s:embedded}

Figure~\ref{f:co2_embodied_all} compares the embodied carbon of SoCs and peripherals, reported in CO\(_2\)-eq to standardize greenhouse gas impacts. Our unit-level model (Equation~\ref{e:co2_unit_embodied}) shows 10\%–20\% deviation from reported values, aligning with prior work~\cite{Gupta:ISCA2022,faiz2024llmcarbon}. By modeling unit-level contributions without altering total chip area, it preserves chip-level accuracy. The detailed breakdowns:
\begin{itemize}
\item \textit{Peripheral Devices}. Embodied carbon estimates include 1.43 kgCO\(_2\)-eq for the camera, 0.04 for the microphone, and 0.08 for the speaker~\cite{Pirson:JCP2021}. The 7-inch TFT LCD dominates peripherals at 10.85 kgCO\(_2\)-eq~\cite{Dell:M2013}.  

\item \textit{Rockchip Edge SoCs}. For rk3588, the PCB (43.5 cm\(^2\), CPA: 0.071 kgCO\(_2\)-eq/cm\(^2\)~\cite{RE:PCB2024}) contributes 3.08 kgCO\(_2\)-eq, and the 8 nm SoC (89 mm\(^2\), CPA: 1.2 kgCO\(_2\)-eq/cm\(^2\)) adds 1.07 kgCO\(_2\)-eq, with its NPU (5\% area) at 0.053 kgCO\(_2\)-eq. LPDDR5 DRAM adds 0.42 kgCO\(_2\)-eq~\cite{Gupta:ISCA2022, Jones:IEDM2023}, with 10.4\% of rk3588’s embodied carbon tied to LLM inference. As a comparison, The PCB, SoC, NPU, and DRAM of rk3568 (fewer CPU cores and a lower-throughput NPU) contribute 2.84, 0.94, 0.03, and 0.38 kgCO\(_2\)-eq, respectively (9.9\% for LLM inference).  

\item \textit{NVIDIA Edge SoCs}. For AGX Orin, the PCB (121 cm\(^2\)) and SoC (455 mm\(^2\)) contribute 8.6 and 5.46 kgCO\(_2\)-eq, with its GPU (35\% area) at 1.91 kgCO\(_2\)-eq and LPDDR5 DRAM at 1.68 kgCO\(_2\)-eq; 22.8\% of its embodied carbon supports LLM inference. AGX Orin’s footprint is $3.44\times$ rk3588’s. As a comparison, for Orin NX (fewer CPU cores, a lower-throughput GPU and a narrower-bandwidth DRAM), PCB, SoC, GPU, and DRAM add 6.4, 2.8, 0.81, and 0.88 kgCO\(_2\)-eq, with 15.4\% linked to LLM inference.
\end{itemize}

\section{Use Cases}

We showcase CO2-Meter through three use cases: (1) analyzing operational energy of an LLM-based edge application, (2) examining operational vs. embodied carbon trade-offs, and (3) assessing LLM inference performance against embodied carbon across edge platforms. These cases highlight CO2-Meter’s role in quantifying and optimizing LLM deployment impacts, focusing on flagship SoCs—NVIDIA AGX Orin and Rockchip rk3588.

\begin{figure}[t!]
\centering
\includegraphics[width=\linewidth]{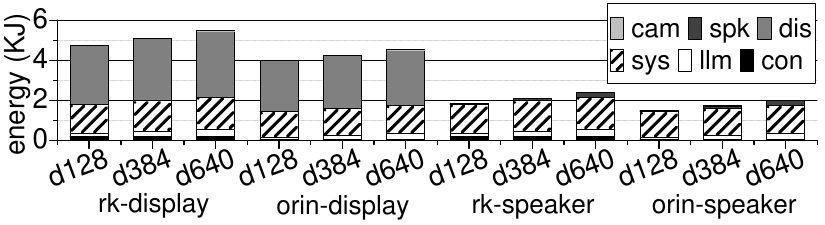}
\vspace{-0.2in}
\caption{The operational energy of the LLM-based virtual reality assistant with a camera input (d$n$: generated token length $n$; cam: camera; spk: speaker; dis: display; sys: system background; llm: LLM; con: conversion).}
\label{f:carbon_app_en2}
\vspace{-0.2in}
\end{figure}

\subsection{Operational Energy Analysis}
\label{s:operational}

Considering the suitable application scenarios of an edge device, we evaluated the energy use of an LLM-based virtual reality assistant~\cite{Min:ISMAR2024} that processes 1.5K-token questions via camera or microphone, converts inputs with image-to-text or voice-to-text, runs inference with Q1.5, and outputs responses through a display or speaker. With microphone input, energy consumption scales with output length (128–640 tokens, Fig.~\ref{f:carbon_app_en1}). The TFT LCD dominates ($>$55\% of total energy), and replacing it with speakers cuts energy usage by over 50\%. Background energy from idle components (e.g., CPU) is the next largest share, suggesting low-power states could save more. LLM inference accounts for only 2–12\% of total energy, so faster NPUs/GPUs can improve efficiency. Despite higher power draw, AGX Orin uses 14–15\% less energy than rk3588 for the same task at all output lengths.

\begin{figure}[t!]
\centering
\includegraphics[width=\linewidth]{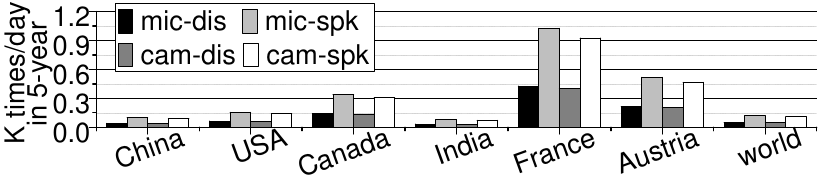}
\vspace{-0.2in}
\caption{The \# of requests for the LLM-based virtual reality assistant making operational and embodied carbon equal (mic: microphone; dis: display; cam: camera; spk: speaker).}
\label{f:carbon_two_ratio}
\vspace{-0.1in}
\end{figure}

Switching to camera input (Fig.~\ref{f:carbon_app_en2}) cuts the virtual reality assistant’s total energy use by 7–21\%, since the camera captures just 2–3 frames instead of recording ~7 minutes of audio. With camera input, AGX Orin uses 17–20\% less energy than rk3588 across all answer lengths.

\subsection{Operational and Embodied Emissions}

AGX Orin exhibits higher embodied carbon but lower operational energy consumption than rk3588, raising the question of how long its energy savings must persist to offset embodied emissions. Unlike data centers strategically placed in low-carbon regions, edge devices depend on the carbon intensity of local electricity grids, which varies widely across countries~\cite{Ritchie:2020}—from below 0.1\,kgCO\(_2\)/kWh in France to about 0.7\,kgCO\(_2\)/kWh in India, with a global average near 0.48\,kgCO\(_2\)/kWh.

Figure~\ref{f:carbon_two_ratio} shows how often the LLM-based virtual reality assistant must be used over a 5-year lifespan for AGX Orin’s operational savings to offset its higher embodied carbon. Low-carbon regions (e.g., France, Austria) require more daily invocations to reach this break-even point. Among input–output settings, the camera–speaker (cam–spk) mode delivers the least operational carbon savings, demanding the highest usage frequency to offset AGX Orin’s embodied carbon versus rk3588.

\begin{figure}[t!]
\centering
\includegraphics[width=\linewidth]{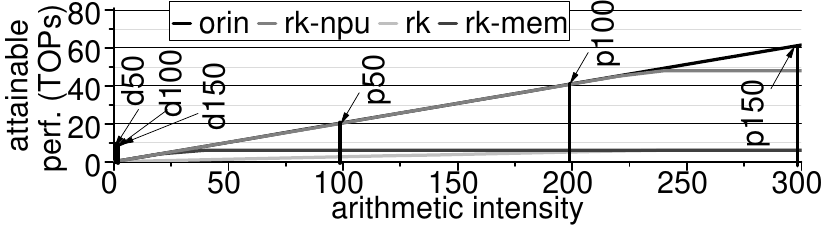}
\vspace{-0.2in}
\caption{The Q1.5 inference Roofline model (perf.:performance; d: decode; p: prefill; the \#s after $d$ and $p$ are prompt lengths; the generated token number is 1; rk: rk3588; orin: AGX Orin; mem: memory; npu: NPU).}
\label{f:co2_roofline_model44}
\vspace{-0.2in}
\end{figure}

\subsection{Performance and Embodied Carbon}

\begin{figure}[t!]
\centering
\includegraphics[width=\linewidth]{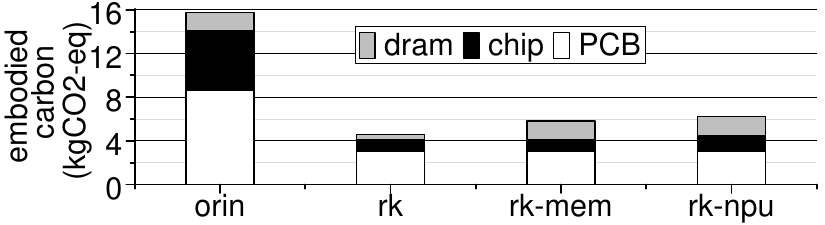}
\vspace{-0.2in}
\caption{Embodied carbon comparison between various hardware configurations (rk: rk3588; orin: AGX Orin; mem: memory; npu: NPU).}
\label{f:co2_roofline_emb}
\vspace{-0.2in}
\end{figure}

Given that embodied carbon often dominates total emissions in edge SoCs, LLM performance upgrades must be weighed against their embodied impact. Key findings include:
\begin{itemize}
\item \textit{Decode Phase Bottleneck}: Edge SoCs like rk3588 and AGX Orin are not optimized for the decode phase of LLM inference due to limited LPDDR bandwidth. As shown in Figure~\ref{f:co2_roofline_model44}, Q1.5’s decode phase sustains low arithmetic intensity ($\sim$2) across prompt lengths (d50–d150), indicating a memory-bound regime. While server GPUs use HBM to overcome this, such solutions are impractical for edge devices with tight power budgets ($\leq$20 W), and viable decode-optimized hardware remains unclear.

\item \textit{Prefill Gains from LPDDR}: Boosting LPDDR bandwidth improves prefill performance. Replacing rk3588’s DRAM with AGX Orin’s LPDDR5 (rk-mem) yields a $2.3\times$ speedup for a 50-token prompt (Figure~\ref{f:co2_roofline_model44}) with only a 27.5\% increase in embodied carbon (Figure~\ref{f:co2_roofline_emb}).

\item \textit{Enhancing Low-End Devices}: On rk3588, both faster memory and more compute are needed. Scaling the NPU $8\times$ and adopting LPDDR5 (rk-npu) delivers $6.8\times$ and $8\times$ speedups for 100- and 150-token prompts (Figure~\ref{f:co2_roofline_model44}), at a 35.7\% embodied carbon cost (Figure~\ref{f:co2_roofline_emb}).
\end{itemize}

\section{Conclusion}

This work introduces \textit{CO2-Meter}, a unified framework for quantifying the end-to-end carbon footprint of LLM inference on edge devices, encompassing both operational and embodied emissions. By modeling peripheral energy via equations, capturing phase-specific inference energy with a GNN-based predictor, and formulating a unit-level embodied carbon model for SoCs, \textit{CO2-Meter} bridges key gaps in existing estimators. Validation confirms its accuracy, and case studies highlight its utility in identifying carbon bottlenecks. \textit{CO2-Meter} enables precise carbon assessment for sustainable LLM deployment, advancing greener hardware–software co-design, carbon-aware policy, and accountability in AI’s environmental impact.

\section{Acknowledgments}
This work was supported in part by NSF CCF-2105972, OAC-2417589, and CAREER AWARD CNS-2143120. 

\bibliography{co2edge}

\end{document}